\documentclass[a4paper,dvips]{article}
\usepackage{icrc2013}
\title{EAS thermal neutron lateral and temporal distributions}

\shorttitle{EAS thermal neutrons}

\authors{
Yu.V.Stenkin$^{1}$, D.M.Gromushkin$^{2}$, A.A.Petrukhin$^{2}$,
 O.B.Shchegolev$^{1}$, V.I. Stepanov$^{1}$, V.I.Volchenko$^{1}$,
I.I. Yashin$^{2}$ and E.A. Zadeba$^{2}$. }

\afiliations{
$^1$ Institute for Nuclear Research of Russian Academy of Sciences, Moscow \\
$^2$ National Research Nuclear University MEPhI, Moscow \\
}

\email{yuri.stenkin@rambler.ru}

\abstract{A novel type of EAS array (PRISMA-32) has been
constructed on the base of NEVOD-DECOR experiment (MEPhI, Moscow)
and is now taking data. It consists of 32 specially designed
scintillator en-detectors able to measure two main EAS components:
hadrons (n) and electrons (e). First results on thermal neutron
lateral as well as temporal distributions are presented. Obtained
exponential neutron lateral distributions are consistent with that
expected for normal hadron production with exponential transverse
momentum distribution.
 As there are no other experimental data on thermal neutron distributions and so,
 to compare results with other measurements, we additionally obtained electron
 lateral distribution function (using the same detectors) and compared it with NKG - function.
 Recorded neutron temporal distributions are very close to that obtained with data of our previous prototypes.
}


\begin{document}
\maketitle

\section{Introduction}

   An idea of a novel type of EAS array proposed for the first time in 2001 [1] was developed
 later [2, 3] to the PRISMA (PRImary Spectrum Measurement Array) project. The main feature
 of the array is the ability to measure the main EAS component - hadrons, through the measurement
 of secondary thermal neutrons produced by these hadrons. Special inorganic scintillator detector
 (en-detector) has been developed for this purpose. The main feature of the detector is
 sensitivity to two EAS components: hadrons - the main one and electrons - the most numerous one.
 A prototype of such array consisting of 32 en-detectors (PRISMA-32) is running now in Moscow
 on a the base of the NEVOD-DECOR experiment (MEPhI). It started in February 2012. Details of
 the en-detector design and of the PRISMA-32 array can be found elsewhere [4, 5].

\section{Data acquisition and pre-analysis}

32 en-detectors array (composed of two 16-detector clusters) is
located inside the experimental hall situated on the 4th floor of
the NEVOD building in MEPhI. Thin layer (~$\sim$30 mg/cm$^{2}$) of
special inorganic scintillator ZnS(Ag) + $^{6}$LiF of 0.36 m$^{2}$
area is placed at the bottom of cylindrical polyethylene (PE)
200-liter tank which is used as the detector housing. A 5"-PMT
(FEU-200) is mounted on the tank lid. Light reflecting cone made
of foiled PE foam of 5-mm thickness is used for better light
collection. As a result, we collect $\sim$50-100 photoelectrons
per a neutron capture. All pulses are integrated with the time of
1 $\mu s$. FADC (ADLINK 10 bit PCI slot PCI-9810) is used for
pulse shape digitizing (20000 samples with a step of 1 $\mu s$).
First pulse produced mostly by EAS electrons is used for energy
deposit measurements and delayed neutron capture pulses are
counted within a time gate of 20 ms to give the number of
neutrons.
  The detector layout is shown in figure 1. Unfortunately, we were
  pressed to use free space around the water pool and this resulted
  in not uniform array structure. Clusters P1 and P2 are not identical
  in shape but they have identical independent data acquisition systems.
   The first level trigger is very simple: a coincidence of any 2 from 16
   detectors in a time gate of 1 $\mu s$ starts all FADCs. On-line program
   analyzes the data and produces second level triggers: M1 in a case
   of coincidences above threshold of $\sim$ 5 m. i. p. (minimum ionizing particle) in the first
   time bin, M2 in a case of total energy deposit more than 50 particles,
   and M3 if the number of recorded neutrons is more than 4. We use 2
   signals from each detector (from the last 12th dynode and from the
   7th dynode) to make dynamic range as wide as possible. All pulses are digitized.
Clocks of both systems are synchronized and the data are combined
off-line using the time gate of 20 ms. As a result, we have data
only in a case when at least 2 detectors were hit with
energy deposit threshold of ~$\sim$ 5 m. i. p. in each cluster (4-fold
coincidence in total), or when energy deposit exceeded 50 m. i. p. in
each cluster, or the number of recorded neutrons exceeded 4 in
each cluster. Note that en-detector counting rate is very low
($\sim$ 0.5 s$^{-1}$) and thus probability of chance coincidence
inside 1 $\mu$s for 1 cluster is $\sim$
0.5$\times$16$\times$10$^{-6}$=8$\times$ 10$^{-6}$. Counting rate
of real triggers of a cluster is ~$\sim$ 700 per day. Therefore,
probability of chance coincidence between the 2 cluster triggers
is $\sim$ 700/86400$\times$0.02=1.4$\times$$10^{-4}$. In addition
to the physical triggers, every 5 min the program generates a soft
trigger M0 to measure and to monitor chance coincidence level for
neutron detection. The latter was found to be 0.34 per event.

\begin{figure*}[!tb]
  \centering
  \includegraphics[width=0.9\textwidth]{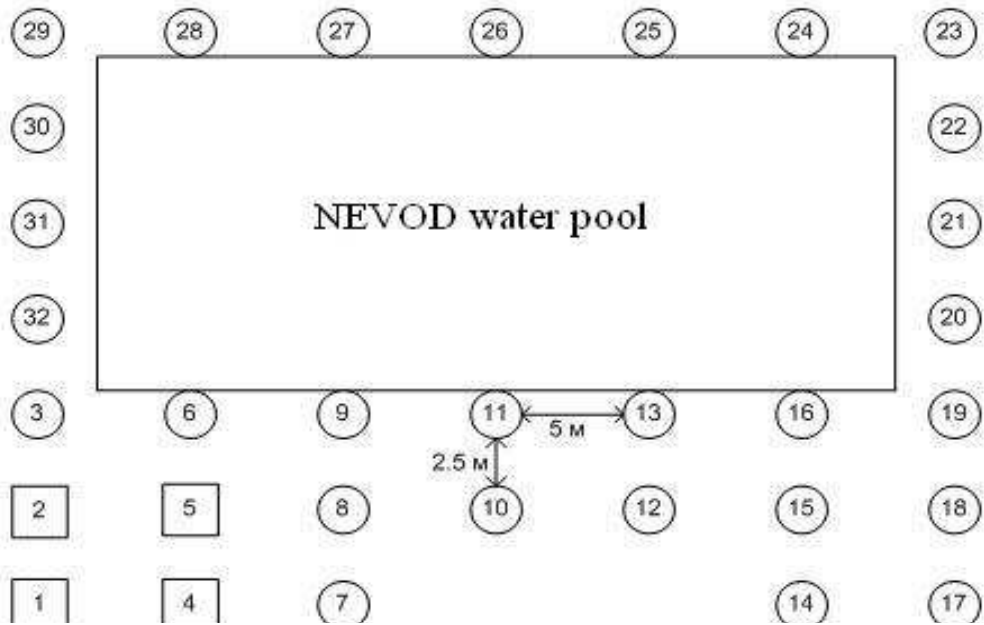}
  \caption{PRISMA-32 lay-out. %
Detectors 1, 2, 4 and 5 are of square shape 0.75 sq. m each;
others - cylindrical 0.36 sq. m. Detectors $1 - 16$ compose
cluster P1, and detectors $17 - 32$ compose cluster P2.}
  \label{fig1}
 \end{figure*}
 \begin{figure}[!tb]
  \centering
  \includegraphics[width=0.45\textwidth]{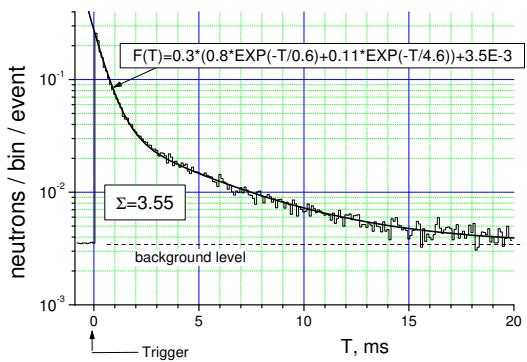}
  \caption{EAS thermal neutron time distributions measured by
  PRISMA-32.  %
T=0 corresponds to the trigger position (EAS passage), $\Sigma$
shows the mean total number of recorded neutrons per event. }
  \label{fig2}
 \end{figure}

\section{Results}

In this report we present experimental data on the "neutron vapor"
parameters obtained up to date. The thermal neutron lateral
distribution in events selected by M1 trigger and 8-fold coincidence with at least 15 m. i. p. in each detector is shown in figure
2. As one can see, the measured time distribution is fitted very
well by the two-exponentional function as it was shown in our
previous works [6, 7]. The first time component is close (but not
equal to) to 1 ms as expected for the thermal neutron lifetime in
rock, soil or concrete. This so-called local neutron component is
produced by high energy hadrons in the concrete floor in a vicinity of the detector. The
second time component could be explained by an admixture of the neutrons produced in the thick concrete roof and neutrons produced in the atmosphere introduced in earlier our works [2-4]. Atmospheric neutrons are produced in air at distances less or about one
hadronic interaction length ($\sim$750 m at sea level) above the
detectors and was expected to have rather large delay. First
measured time parameter  less than it could be expected. Why?
In our opinion, the first parameter is equal to only 0.6 ms due to
the influence of the NEVOD water pool. It is known that neutron
lifetime in water is $\sim$0.2 $ms$, i.e. much less than in soil.
That is why a presence of water must decrease the lifetime of
recorded neutrons. As for the second time parameter, it is much less then it is expected for atmospheric neutrons. The PRISMA-32 array is located inside the building having rather
thick concrete walls and roof (from 20 to 50 cm). Therefore, it is
difficult for atmospheric thermal neutrons to penetrate the
experimental hall. On the other hand, the roof and the walls serve
as an additional targets for hadron interactions and as a
moderator and a thermalizer for fast neutrons. As a result,
thermal neutrons from the roof and the walls must have an
additional delay to reach the detector ($\sim$0.5 ms/m). Taking
into account that the height of the roof is $\sim$7-8 m above the
detectors, one could expect the delay of $\sim$ 4 ms. So, in our
opinion, the second observed time parameter is probably explained mainly
by the concrete roof and walls.

 \begin{figure}[!tb]
  \centering
  \includegraphics[width=0.45\textwidth]{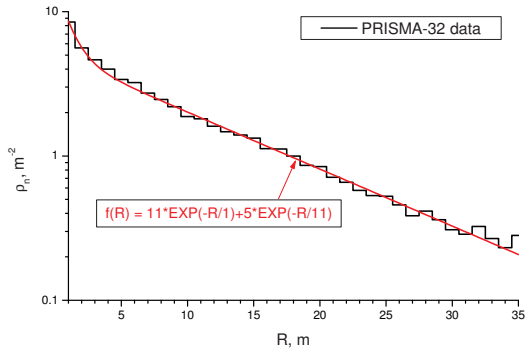}
  \caption{EAS thermal neutron lateral distribution. R - distance from the EAS core.}
  \label{fig3}
 \end{figure}

\begin{figure}[!tb]
  \centering
  \includegraphics[width=0.45\textwidth]{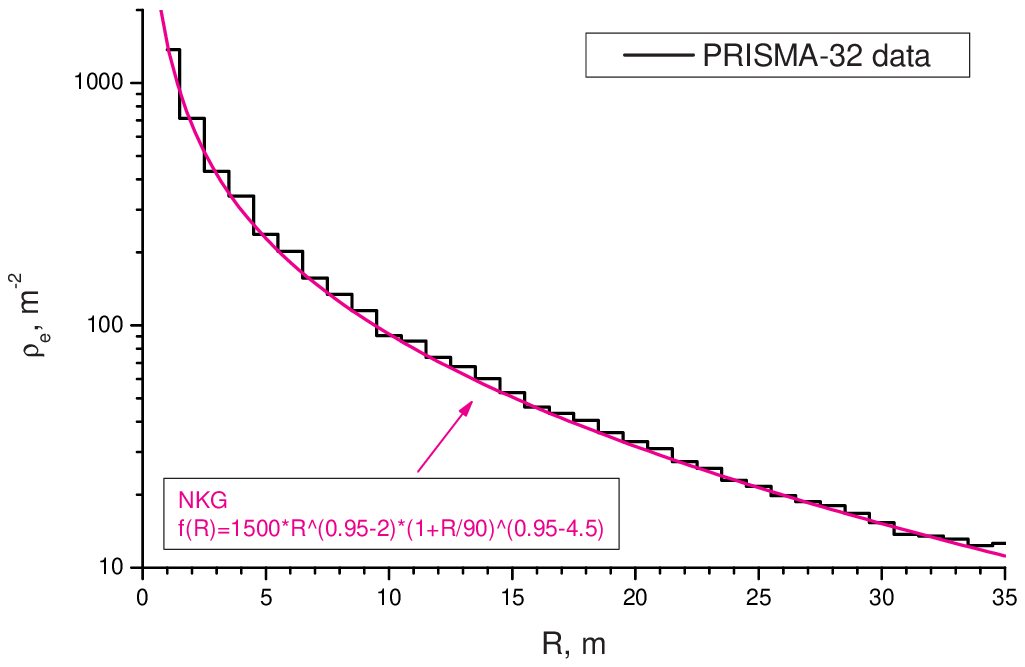}
   \includegraphics[width=0.45\textwidth]{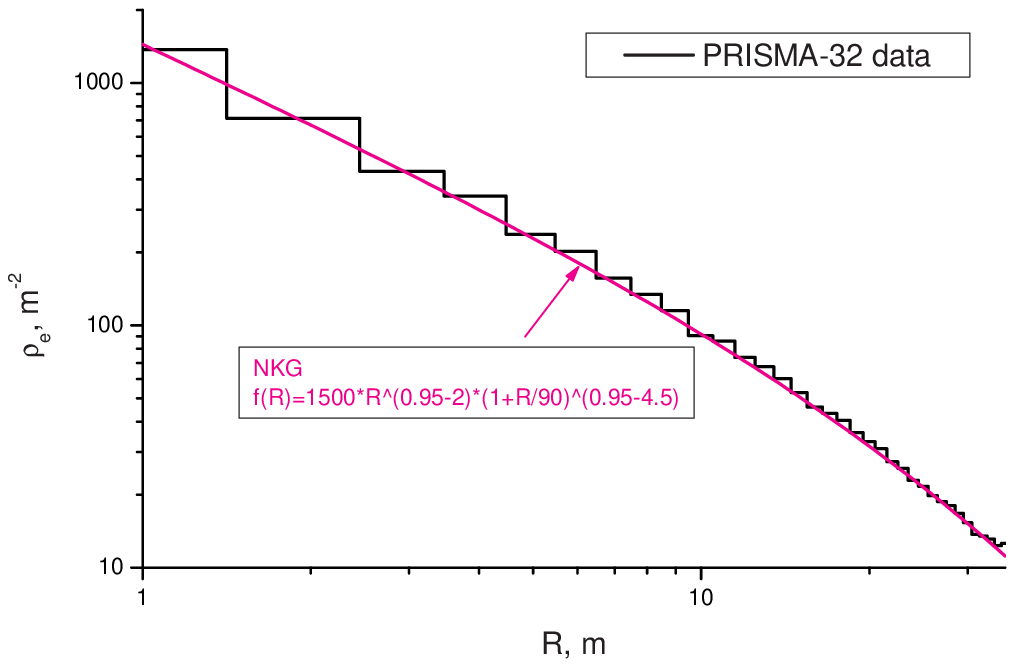}
  \caption{Electron lateral distribution measured by the PRISMA-32 with the
  en-detectors. %
  Upper panel: Linear - Log scale; Lower panel: Log - Log scale.
  }
  \label{fig4}
 \end{figure}

EAS thermal neutron lateral distribution is presented in figure 3
along with a fitting function. Here again double exponential
function fits the data rather well. The first distance parameter
(1 m) is probably connected with the characteristic distance
between recorded neutron and its parent hadron. In other words,
recorded thermal neutrons produced locally (local neutrons) are
collected from the nearest vicinity of the detector. Keeping this
in the mind one can conclude that the second distance parameter
(11 m) is connected with the EAS hadron lateral distribution
following the exponential distribution in transverse momentum.
    It is obvious that a hadron with momentum $P_{h}$ produced at one interaction
     length above the observational level ($\Delta$H) with a transverse momentum
     $P_{t}$ (note that $<P_{t}> \approx 0.4 GeV/c$)
     could be found at a distance R from the EAS axis:
     \begin{equation}
         R =\Delta H \times P_{t} /P_{h}
          \label{1}
     \end{equation}
It is also possible to estimate from (1) the mean energy of
hadrons (mean momentum) producing the recorded neutrons as: $<
P_{h}>$ $\approx$ $\Delta$H $\times$ $<P_{t}> / R$ =750$\times$
0.4/11 $\approx$ 27 GeV/c.
    To compare our results with other measurements, we additionally obtained
    electron lateral distribution function (using the same en-detectors) and compared
    it with NKG - function. Experimental results along with fitting functions are shown
    in figure 4 in linear (left panel) and logarithmic (right panel) R-axis scale. These data also were selected by M1 with 8-fold coincidence with 15 m. i. p. threshold.
     It is seen that NKG power law function can fits the data very good but
     with the abnormally small age parameter s = 0.95. Probably
     this is also due to the roof presence.

\section{Conclusion}

A novel method of EAS study has been developed, detector prototype
PRISMA-32) has been realized and operates now. The measured
parameters of EAS neutrons time and lateral distributions are
close but not equal to our initial expectations. Location of the
array inside the building does not probably allow us to record
atmospheric neutrons, i. e. neutrons born in the atmosphere.
Measured electron lateral distribution at distances 1-30 m seems
to be fitted good with the NKG-function with age close to 1. Results of this work and [8] will help us to optimize
the future full-scale PRISMA array's design.

\vspace{0.5cm}
\footnotesize{{\bf Acknowledgment:} {The work is fulfilled in
Scientific and Educational Center NEVOD. Authors would like to
acknowledge the financial support from RFBR (grant 11-02-01479),
from the RAS Presidium Program " Fundamental properties of matter
and Astrophysics", and Ministry of Education and Science of RF. }}

\end{document}